\documentclass[aps, prd, twocolumn, superscriptaddress, nofootinbib, showpacs,
floatfix]{revtex4}
\usepackage[dvips]{epsfig}

\usepackage{natbib}

\def\lsim{\lower0.6ex\vbox{\hbox{$ \buildrel{\textstyle <}\over{\sim}\ $}}}
\def\gsim{\lower0.6ex\vbox{\hbox{$ \buildrel{\textstyle >}\over{\sim}\ $}}}

\def\beq{\begin{equation}}
\def\eeq{\end{equation}}

\def\beqa{\begin{eqnarray}}
\def\eeqa{\end{eqnarray}}

\def\bfig{\begin{figure}[ht] \begin{center}}
\def\bfigh{\begin{figure}[h!] \begin{center}}
\def\bfigb{\begin{figure}[hb!] \begin{center}}
\def\bfigt{\begin{figure}[t!] \begin{center}}
\def\bfight{\begin{figure}[ht!] \begin{center}}
\def\efig{\end{center} \end{figure}}

\def\btab{\begin{table*}[ht]}
\def\etab{\end{table*}}

\def\apjs{Astrophys.\ J.\ Supp.}
\def\apj{Astrophys.\ J.}

\def\aap{Astron.\ \& Astrophys.}
\def\mnras{Mon. Not. Roy. Astron. Soc.}

\begin{document}

\title{Direct X-ray Constraints on Sterile Neutrino Warm Dark Matter}

\author{Casey R. Watson}
\email{cwatson@mps.ohio-state.edu}
\affiliation{Department of Physics, The Ohio State University,
Columbus, OH 43210, USA}
\author{John F. Beacom}
\affiliation{Department of Physics, The Ohio State University, 
Columbus, OH 43210, USA}
\affiliation{Department of Astronomy, The Ohio State University,
Columbus, OH 43210, USA}
\author{Hasan Y\"uksel}
\affiliation{Department of Physics, The Ohio State University, 
Columbus, OH 43210, USA}
\author{Terry P. Walker}
\affiliation{Department of Physics, The Ohio State University, 
Columbus, OH 43210, USA}
\affiliation{Department of Astronomy, The Ohio State University,
Columbus, OH 43210, USA}


\date{18 May 2006}

\begin{abstract}
Warm dark matter (WDM) 
might more easily account for small scale clustering measurements 
than the heavier particles typically invoked in 
$\Lambda$ cold dark matter ($\Lambda$CDM) cosmologies.  In this paper,
we consider a $\Lambda$WDM cosmology in which sterile neutrinos 
$\nu_s$, with a mass $m_s$ of roughly 1-100 keV, are the dark matter. 
We use the diffuse X-ray spectrum
(total minus resolved point source emission) 
of the Andromeda galaxy
to constrain the rate of sterile neutrino radiative decay:
$\nu_s \rightarrow \nu_{\rm e,\mu ,\tau} + \gamma$.
Our findings demand that $m_s < 3.5 \rm\ keV$ (95\% C.L.) which is a significant improvement
over the previous (95\% C.L.) limits inferred from the X-ray emission of nearby clusters,
$m_s < 8.2$ keV (Virgo A) and $m_s < 6.3$ keV (Virgo A + Coma). 
\end{abstract}

\maketitle

\section{Introduction}

Standard $\Lambda$CDM models of 
galaxy formation predict more small scale structure than is observed, e.g.,
excess galactic satellites (``the missing satellite problem''
\cite{Kauffmann:1993gv,Klypin:1999uc,Moore:1999wf,Willman:2004xc}),
high galactic central densities (``the central density problem''
\cite{Dalcanton:2000hn,vandenBosch:2000rz,Zentner:2002xt,Zentner2003,Strigari2006}), etc.
Because structure formation 
is suppressed on scales smaller than
$\lambda_{FS} \simeq 0.25_{~}\rm{Mpc} \left(m_{\rm DM}/\rm~keV \right)^{-4/3}$
for dark matter particles of mass $m_{\rm DM}$ \cite{Pad}, 
models of Warm Dark Matter (WDM) with $\sim$ keV masses,
such as sterile neutrinos \cite{Dodelson1994,AFP,Dolgov2002}, 
may lead to better agreement with observations 
than standard (GeV -- TeV) CDM candidates, like the neutralino.
On the other hand, as much as the suppression of small scale structure by WDM may alleviate
problems at low redshifts, the same suppression will also delay the onset of reionization at higher redshifts \cite{Barkana2001,Yoshida2003},
possibly to an extent that is difficult or impossible to reconcile with the WMAP3 results \cite{Spergel2006}.
Constraints on the level of small scale clustering from measurements of the CMB
\cite{hinshaw03,kogut03,readhead04,Jones:2005yb,Kuo:2002ua,dickinson04}, 
the 3-D galaxy power spectrum \cite{Tegmark:2003uf}, and the Lyman-$\alpha$ forest
\cite{Narayanan:2000tp,McDonald:2004eu,McDonald:2004xn,Croft:2000hs,
Viel:2004bf,Viel:2004np,Viel:2005qj}
also preclude drastic small scale suppression.
In the context of the sterile neutrino ($\nu_s$)
WDM model presented in 
\cite{AFP,AFT,Aba05a,Aba05b}, 
these data require $m_s > 1.7 \rm~keV$ (95\% C.L.) according to the
analysis presented in \cite{Aba05b}. 
More recent work by Seljak et al. \cite{Seljak} suggests that this lower bound 
can be improved by almost an order of magnitude to
$m_s > 14 \rm~keV$ (95\% C.L.). 
Because indirect limits imposed by 
small scale clustering data require the interpretation of
simulations at their resolution limit and because of the striking improvement of this constraint over previous work, 
this result has already generated some debate \cite{Aba05b, AbaSav}.
Even more recently, however, Viel et al. \cite{Viel2006} reported a very similar indirect limit
of $m_s > 10 \rm~keV$ (95\% C.L.), essentially confirming the results of Ref. \cite{Seljak}.

Fortunately, it is also possible to directly constrain $m_s$ 
based on the 
radiative decay of sterile neutrinos to 
X-rays of energy $E_{\gamma, \rm s} = m_s/2$ via
$\nu_s \rightarrow \nu_{\rm e,\mu ,\tau} + \gamma$.
This test is \it{direct}\rm~in the sense that it probes the signature of 
individual particle decays, whereas 
the cosmological tests are \it{indirect}\rm~in the sense that 
they probe only the macroscopic clustering
signatures of the dark matter. In order to prevent a potentially viable
dark matter candidate from being dismissed prematurely, it is important
to separately improve both constraints.  This is especially true in the
case of sterile neutrinos which, apart from their possible role as the dark matter,
have interesting implications for several important physical processes including
the production of the baryon \cite{AS2005,ABS2005,AKS2006,RHP2006} and lepton \cite{ABFW}
asymmetries, Big Bang Nucleosynthesis \cite{Dolgov2000,ABFW}, reionization
\cite{Hansen2004,Kusenko_Reion,OShea2006,Mapelli_Reion}, 
neutrino oscillations \cite{Gelmini2004,Cirelli2005,Smirnov2006},
pulsar kicks \cite{Kusenko1998,Fuller2003,Kusenko2004,Barkovich2004,Fryer2005}, etc.

Several direct limits on $m_s$ have already been reported.
The absence of anomalous line features in
\it{XMM-Newton}\rm~(0.5$-$12 keV) \cite{Lumb,ReadPonX}
and HEAO-I (3$-$60 keV) \cite{Marshall,Gruber} measurements of the
Cosmic X-ray Background (CXB) allowed Boyarsky et al. \cite{Boyarsky:2005us}   
to set a (95\% C.L.) upper limit of $m_s  < 9.3 \rm~keV$ (assuming, as with all the mass
limits quoted below, that $\Omega_{DM} = \Omega_s = 0.24$).
In Ref.~\cite{AFT}, Abazajian, Fuller, and Tucker (hereafter AFT)
suggested that the best constraints
could be achieved by examining individual objects, e.g., galaxies and/or clusters,
rather than the CXB.
Based on \it{XMM-Newton}\rm~observations of Virgo A (M87), the dominant galaxy in the northern part of the
Virgo cluster \cite{VirgoX},
Abazajian \cite{Aba05b} arrived at an improved upper limit of
$m_s  < 8.2 \rm~keV$ (95\% C.L.).
Boyarsky et al. \cite{Boyarsky2006} have also used \it{XMM-Newton}\rm~observations 
of Virgo A \cite{VirgoX} and the Coma \cite{Briel2001,Neumann2003} 
cluster to explore possible constraints on $m_s$, but
they do not provide a definite numerical limit.
Abazajian and Koushiappas \cite{AbaSav} estimate
an upper bound of $m_s  < 6.3 \rm~keV$ (95\% C.L.)
based on the results in Ref.~\cite{Boyarsky2006}. 

An important general feature of all sterile neutrino mass constraints is that $m_s$ is very
weakly dependent on the flux from which it is
inferred: $m_s \propto \Phi_{\rm x,s}^{0.3}$ \cite{Aba05a,Aba05b} 
(Eqn.~\ref{Phixs_nomix} below). Consequently, even a significantly improved
constraint on $\Phi_{\rm x,s}$
leads to only a modest improvement in the mass limit.  
This dependence puts a premium on carefully selecting an object
with the best ratio of dark matter decay signal to astrophysical background.

In this paper, we consider the limits imposed by \it{XMM-Newton}\rm~observations of the
Andromeda galaxy (M31) \cite{M31}.
Although these observations probe significantly less dark matter
than observations of clusters, 
the predicted sterile neutrino decay signal from Andromeda
is comparable to that of Virgo A 
due to Andromeda's close proximity ($D_{\rm M31} \simeq$ $0.78 \pm 0.02$ Mpc \cite{DM31}, as compared to
$D_{\rm M87} \simeq 15.8 \pm 0.8$ Mpc \cite{DM87}).  
Moreover, by analyzing a single galaxy rather than a cluster, 
we avoid one substantial source of astrophysical background: hot,
intra-cluster gas. Shirey et al. \cite{M31} eliminate a second major 
astrophysical background by identifying 
and removing the flux from resolved X-ray point sources in Andromeda. 
The only remaining astrophysical contributions
to the diffuse spectrum of Andromeda are
unresolved point sources
and (a modest amount of) hot gas.
The analysis of Shirey et al. \cite{M31} suggests that hot gas dominates the diffuse emission
at low energies but falls off rapidly at $E_\gamma ~\gsim$ 0.8 keV
until the unresolved point sources begin
to dominate at $E_\gamma ~\gsim$ 2 keV. 
This fact underscores the importance of using high resolution spectra to constain $m_s$,
without which we could not benefit from the rapidly decreasing astrophysical emission
in the energy range of interest ($E_\gamma ~\gsim$ 0.85 keV) and would, instead,
be forced to compare $\nu_s$ decay peaks to much larger
\it{broad-band}\rm~(e.g., soft band: 0.5-2 keV) fluxes. 
For these reasons, the diffuse spectrum of Andromeda is an excellent data set
for constraining the decay of sterile neutrinos and, in particular,
demands $m_s  < 3.5 \rm~keV$ (95\% C.L.), as we show below. 

In Sec.~\ref{Model}, we provide a brief overview of the sterile neutrino WDM model \cite{AFP,AFT}.
In Sec.~\ref{Galsamp}, we describe the \it{XMM-Newton}\rm~observations of Andromeda \cite{M31}
and Virgo A \cite{VirgoX}, and other relevant physical
properties of these systems (see Table I). 
In Sec.~\ref{GalClus}, we explain our analysis
and compare the potential for direct detection of keV dark
matter decay in nearby galaxies and clusters. 
In Sec.~\ref{mslimit} we present an updated sterile neutrino
exclusion plot, including our mass limit
based on the diffuse spectrum of Andromeda as well as the
previous direct and indirect limits.
We conclude in Sec.~\ref{conclusions}.
Throughout the paper, we assume a flat cosmology 
with $\Omega_{\rm baryon}=0.04$, $\Omega_{\rm WDM}= \Omega_{\rm s}= 0.24$,
$\Omega_{\Lambda}=0.72$, and $h = H_0/100~ \hbox{km s}^{-1} ~\rm  Mpc^{-1} = 0.72$.

\section{The Sterile Neutrino WDM Model}
\label{Model}

In the model presented in Refs.$_{~}$\cite{AFP,AFT,Aba05a,Aba05b},
sterile neutrinos $\nu_s$ with $m_s < m_e$ predominantly decay to three light, active neutrinos
$\nu_\alpha$ (i.e., $\nu_{\rm e,\mu ,\tau}$) 
at a rate of 
\beq
\Gamma_{3\nu} \simeq 8.7\times 10^{-31} \rm{s}^{-1} 
\left(\frac{\sin ^{2}2\theta}{10^{-10}}\right)
\left(\frac{m_s}{\rm~keV}\right)^5,
\eeq
where $\theta$ is the vacuum mixing angle for 
an effective 2$\times$2 unitary transformation from $\nu_s \leftrightarrow \nu_\alpha$;
since we consider only very small mixing angles, $\nu_s$ is nearly a pure mass eigenstate
\cite{AFP,AFT}. The radiative decay in which we are interested, 
$\nu_s \rightarrow \nu_{\alpha} + \gamma$,
is suppressed by a factor of
$ 27\alpha /8\pi \simeq 1/128$ \cite{Barger1995,Drees2000} relative to $\Gamma_{3\nu}$ and occurs at a rate of
\cite{AFP,AFT} 
\beq
\Gamma_s \simeq 6.8\times 10^{-33} \rm{s}^{-1} 
\left(\frac{\sin ^{2}2\theta}{10^{-10}}\right)
\left(\frac{m_s}{\rm~keV}\right)^5.
\eeq
While the decay rate is very slow, a large collection of sterile neutrinos will produce a detectable
X-ray signal. In particular,
the X-ray luminosity resulting from the decay of $N_s = (M_{\rm DM}/m_s )$ sterile neutrinos
in a dark matter halo of mass $M_{\rm DM}$ is given by
\beqa
L_{\rm x,s} = E_{\gamma ,\rm s}N_{\rm s}\Gamma_s  
    = \frac{m_s}{2}\left(\frac{M_{\rm DM}}{m_s}\right)\Gamma_s \nonumber \\
    \simeq 6.1 \times 10^{32} \rm{erg~s}^{-1} 
    \left(\frac{M_{\rm DM}}{10^{11}M_\odot}\right)\\ \nonumber
    \times \left(\frac{\sin ^{2}2\theta}{10^{-10}}\right) 
    \left(\frac{m_s}{\rm~keV}\right)^5.
\label{Lxs}
\eeqa
The corresponding line flux at $ E_{\gamma ,\rm s} = m_s/2$ is
\beqa
\Phi_{\rm x,s} \simeq 5.1 \times 10^{-18} \rm{erg~cm}^{-2}\rm{s}^{-1} 
    \left(\frac{D}{\rm Mpc}\right)^{-2} \nonumber \\
    \times \left(\frac{M_{\rm DM}}{10^{11}M_\odot}\right)  
    \left(\frac{\sin ^{2}2\theta}{10^{-10}}\right) 
    \left(\frac{m_s}{\rm~keV}\right)^5.
\label{Phixs}
\eeqa
If we assume a QCD phase-transition
temperature of $T_{\rm QCD} = 170$ MeV 
and a lepton asymmetry of $L \simeq \eta_{10} \equiv n_{\rm baryon}/n_\gamma \simeq 10^{-10}$) \cite{Aba05a, AbaSav},
the sterile neutrino density-production 
relationship \cite{Aba05a} (updating \cite{AFT}) is
\beq
m_s = 3.27\rm~{keV}\left(\frac{sin^{2}2\theta}{10^{-8}}\right)^{-0.615}
\left(\frac{\Omega_{\rm s}}{0.24}\right)^{0.5}. 
\label{Omega_sin2th}
\eeq
Although the values of $T_{\rm QCD}$ and $L$ are uncertain, we adopt Eqn.~(\ref{Omega_sin2th})
for definiteness and for ease of comparison to the literature.  
Our limit on $m_s$ could easily be re-evaluated
for a different density-production relationship; for example, note
the three lines for models with large $L$ in Fig.~\ref{XMM_ms} below.
By combining Eqns.~(\ref{Phixs}) and (\ref{Omega_sin2th}),
we arrive at an expression for the line flux that is independent of 
the mixing angle:
\beqa
\Phi_{\rm x,s}(\Omega_{\rm s}) \simeq 3.5 \times 10^{-15} \rm{erg~cm}^{-2}\rm{s}^{-1} 
    \left(\frac{D}{\rm Mpc}\right)^{-2} \nonumber \\
    \times \left(\frac{M_{\rm DM}}{10^{11}M_\odot}\right) 
    \left(\frac{\Omega_{\rm s}}{0.24}\right)^{0.813}
    \left(\frac{m_s}{\rm~keV}\right)^{3.374}.
\label{Phixs_nomix}
\eeqa
In the remaining sections we describe the data and methods we use to constrain $m_s$
in the context of this model.
 
\section{Properties of Andromeda and Virgo A}
\label{Galsamp}

Comparison of the X-ray emission from Andromeda and Virgo A
is facilitated by the fact that both systems were observed with the same
instrument: \it{XMM-Newton}\rm~\cite{M31,VirgoX}. Even more important 
is the fact that \it{XMM}\rm~is sensitive
to the full (95\% C.L) range of presently allowed $\nu_s$ radiative
decay energies (0.85 keV $< E_{\gamma ,\rm s} = m_s/2 < 3.15$ keV) \cite{AbaSav}.
In addition to their X-ray properties, this section also provides the distances
and dark matter mass estimates we adopt for our calculations of the sterile neutrino
decay fluxes from each object.

\subsection{X-ray Data}

Shirey et al. \cite{M31} observed Andromeda with \it{XMM}\rm~for 34.8 ks 
out to a radius of 15 arcminutes ($\simeq$ 3.4 kpc at $D_{\rm M31} \simeq$ $0.78 \pm 0.02$ Mpc \cite{DM31}).  
The (0.5$-$12 keV) diffuse spectrum we utilize was extracted from within a radius of 5$'$
($\simeq$ 1.1 kpc) from the center of the galaxy.  
Shirey et al. \cite{M31} produce the diffuse spectrum by identifying 
and removing the flux from resolved X-ray point sources in Andromeda; we make
no attempt to model or subtract the remaining astrophysical contributions
from unresolved point sources and hot gas.
We note that the Shirey et al. data are consistent
with the (0.5$-$7 keV) diffuse spectra presented in a joint \it{Chandra$-$XMM}\rm~study
of Andromeda by Takahashi et al. \cite{Takahashi2004}.

B\"ohringer et al. \cite{VirgoX} used \it{XMM}\rm~to measure the flux of and around Virgo A (M87)
out to a radius of 12$'$ 
($\simeq$ 55 kpc at $D_{\rm M87} \simeq 15.8 \pm 0.8$ Mpc \cite{DM87}) over a (usable) exposure time of 25.9 ks.
Abazajian et al. \cite{AFT,Aba05b} consider the (0.5$-$8 keV) flux from within a 
radius of 8.5$'$ ($\simeq$ 39 kpc).  These data are summarized in Table I.

\subsection{Dark Matter Masses}
\label{DMest}

To estimate the dark matter mass $M^{\rm fov}_{\rm DM}$ of Andromeda and Virgo A
that is enclosed within the \it{XMM}\rm~field of view (fov), 
we integrate the ($r^{-2}-$weighted) dark matter density of each halo
$\rho_{\rm DM}(|\vec{r}-\vec{D}|)$ 
over a truncated cone of radius $R_{\rm fov}$ and length $2R_{\rm vir}$:
\beq
\Sigma_{\rm fov} = \int \frac{\rho_{\rm DM}(|\vec{r}-\vec{D}|)dV_{\rm fov}}{ r^{2}}.
\eeq
We define $ M^{\rm fov}_{\rm DM} = {D^2}\Sigma_{\rm fov} $.
The distance to the center of each object is $D$,
$R_{\rm vir}$ is the virial radius,
\beq
R_{\rm fov} = 
\theta_{\rm fov}r 
\simeq 0.3~\rm kpc  
\left(\frac{\theta_{\rm fov}}{1'}\right) 
\left(\frac{r}{Mpc} \right)
\label{fov}
\eeq
is the radial extent of the fov
at a distance $r$, and $r$ varies between $D_{\rm} - R_{\rm vir}$,
at the near ``edge'' of each dark matter halo, to $D_{\rm} + R_{\rm vir}$ at the far ``edge''. 

Based on rotation curve data, Klypin, Zhao, and Somerville \cite{KZS}
estimated the dark matter mass distribution of Andromeda,
$\rho_{\rm DM, M31}$. 
When we integrate $\rho_{\rm DM, M31}$ over $V_{\rm fov, M31}$
($\theta_{\rm fov} = 5'$ \cite{M31}; $R_{\rm vir} \simeq 300$ kpc \cite{KZS}), we find that the 
region from which the diffuse emission spectrum of Andromeda was extracted contains
\beq
M^{\rm fov}_{\rm DM,M31} \simeq (0.13 \pm 0.02) \times 10^{11} M_\odot.
\eeq 
As shown in Ref.~\cite{KZS}, about half of this mass is enclosed within a sphere of 1 kpc radius about the center of Andromeda. 
It is worth re-emphasizing that our mass limit scales like
$\Phi^{0.3}_{\rm x,s} \propto (M^{\rm fov}_{\rm DM})^{0.3}$, and is therefore very insensitive
to uncertainties in $M^{\rm fov}_{\rm DM}$.
To be conservative, we ignore the contribution from the fraction of the Milky
Way halo within the fov; see Refs.~\cite{RHP2006,Boyarsky2006b} for discussion of possible
constraints based on Milky Way dark matter alone, i.e., ``blank sky observations."

In Ref.~\cite{AFT}, AFT integrate an isothermal $\beta$-model \cite{HMS}
to estimate the dark matter mass of Virgo A within a 17$' \times 17 '$ rectangular prism.
When we integrate the same model over a truncated cone of cross-sectional radius
$\theta_{\rm fov}=8.5'$ and length $2R_{\rm vir} \simeq 3.6$ Mpc \cite{HMS}, we find a mass of 
\beq
M^{\rm fov}_{\rm DM,M87} \simeq (0.75 \pm 0.08) \times 10^{13} M_\odot,
\eeq
which agrees with the result of AFT ($M^{\rm fov}_{\rm DM,M87} \simeq 10^{13} M_\odot$)
to within a factor of $V_{\rm fov, cone}/V_{\rm fov, prism} \simeq \pi/4$.

\section{The Detectability of $\nu_s$ Decays: Galaxies vs. Clusters}
\label{GalClus}

\begin{figure}
\includegraphics[height = .35\textheight, width = .5\textwidth]{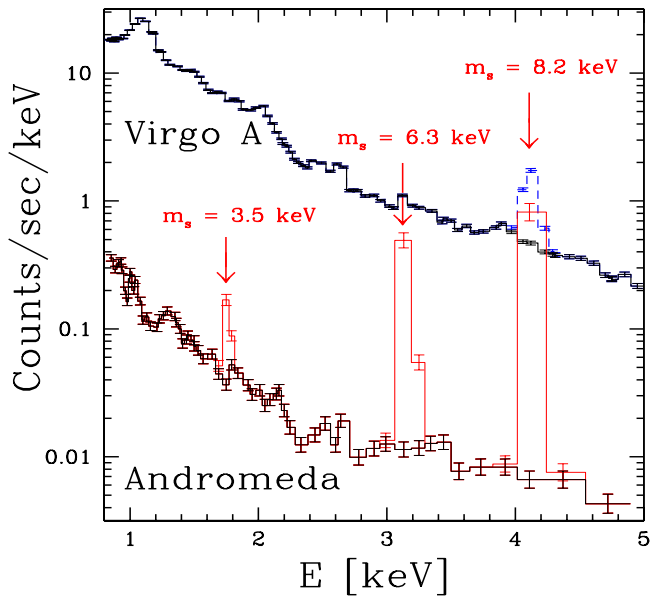}
\caption {Here we compare the detectability of
$\nu_s$ decays in Andromeda \cite{M31} and Virgo A \cite{Aba05b,VirgoX}. 
The first statistically significant ($4\sigma_{\rm f}$) $\nu_s$ decay
peak relative to the measured spectrum of Andromeda occurs at
$E_{\gamma, \rm s} = m_{\rm s, lim}/2$ = 1.75 keV, which excludes $m_s > 3.5$ keV (95\% C.L.). 
According to the analysis
presented in \cite{Aba05b}, the spectrum of Virgo A
excludes $m_s > 8.2$ keV (95\% C.L.), which would produce a decay signature like the dashed histogram. 
Because Andromeda would produce a similar $\nu_s$ decay signal to Virgo A (Eqn.~\ref{DMflux}), but
over a much smaller background,
the prospective decay signature of 8.2 keV sterile neutrinos in Andromeda is enormous by comparison.
As an intermediate case, we also show what the decay peak associated with a 6.3 keV sterile
neutrino, the estimated Virgo A + Coma mass limit \cite{Boyarsky2006,AbaSav}, would look like in Andromeda.
The vertical ($1\sigma$) error bars reflect the Poisson statistics of  
the signal and background count rates measured during each observation.
\label{Galspec}}
\end{figure}

In Fig.~\ref{Galspec}, we compare the diffuse spectrum of Andromeda to the
spectrum of Virgo A and determine the
$\nu_s$ decay signals (in Counts/sec/keV) that would be produced by each object for selected
values of $m_s$.
To calculate the $\nu_s$ decay fluxes, we 
assume that sterile neutrinos comprise all of the dark matter, i.e.,
$\Omega_{s} = \Omega_{\rm DM} = 0.24$,
and evaluate Eqn.~(\ref{Phixs_nomix}) based on the distances and dark matter
masses given in Table I for each object.
Based on the total count rates and flux measurements of Andromeda \cite{M31} and Virgo A \cite{VirgoX},
we divide by factors of $C^{\rm M31}_{\rm x,Ct} = 6.3\times 10^{-12}~\rm{erg}~\rm{cm}^{-2}~\rm{Ct}^{-1}$
and $C^{\rm M87}_{\rm x,Ct} = 7.0\times 10^{-12}~\rm{erg}~\rm{cm}^{-2}~\rm{Ct}^{-1}$
to convert the sterile neutrino fluxes (Eqn.~\ref{Phixs_nomix}) to count rates. 
To realistically simulate the detected ``line'' fluxes in Fig.~\ref{Galspec},
we use a Gaussian centered at 
$E_{\gamma , s} = m_s/2$ with a FWHM of $\Delta E = E_{\gamma , s}/30$ (a conservative
estimate of the energy resolution of the \it{XMM}\rm~EPIC detector\footnote{http://heasarc.nasa.gov/docs/xmm/xmm.html}). Doing so distributes
$\simeq 72\%$ of the signal
over an energy range of $\simeq \Delta E$
and converts the $\nu_s$ decay count rates to the same units as those of the measured spectra, Counts/sec/keV. 

\begin{table}[t!]
\caption{\label{tab:NG} Here we summarize the properties of Andromeda \cite{M31,DM31,KZS} and
Virgo A \cite{Aba05b,VirgoX,DM87,HMS}.  
In rows (2) - (6), we show, respectively, the distance to each object,
the angular radius $\theta_{\rm fov}$ of the \it{XMM-Newton}\rm~field of view (fov) of
each observation, our estimates of the dark matter masses 
probed within each fov, the \it{XMM}\rm~exposure times (in kiloseconds: ks), 
and the (95\% C.L.) upper bounds on $m_s$.  
}
\begin{ruledtabular}
\begin{tabular}{ccc}
Galaxy Name&                                  Andromeda (M31)&    Virgo A (M87)\\ \hline 
Distance (Mpc)&                          $0.78 \pm 0.02$&    $15.8 \pm 0.8$\\ \hline
$\theta_{\rm fov}$ (arcminutes)&                        5.0$'$&  8.5$'$\\ \hline
$M^{\rm fov}_{\rm DM}/10^{11} M_\odot$&  $0.13 \pm 0.02$&   $75 \pm 8$\\ \hline
$t_{\rm exp}$ (ks)&                   34.8&  25.9\\ \hline
\textbf{$m_s$} (keV) (95\% C.L.)&    \textbf{3.5}&  \textbf{8.2}\\
\end{tabular}
\end{ruledtabular}
\end{table}

To determine the mass limit imposed by the diffuse spectrum of Andromeda,
we evaluate the $\nu_s$ decay signal as described above for increasing
values of $m_s$ until we reach the 
first statistically significant ($\geq~ 2\sigma_{\rm f}$) departure from the measured
\it{XMM}\rm~spectrum. To be conservative, we use
the largest fluctuations in the spectral data relative to a smooth (power-law) fit to 
define the statistical significance ($1\sigma_{\rm f}$) of the $\nu_s$ decay signal.
The resulting $2\sigma_{\rm f}$ limit is much more significant than the 95\% C.L.
defined by the formal statistical errors on the measured points.  However, 
because features in the spectra may already reflect sterile neutrino decay and/or 
atomic line emission, detector backgrounds, etc.$_{~}$and are generally of uncertain origin
(at least in the absence of detailed modeling),
we argue that they are the appropriate gauge of statistical significance; i.e., 
for an upper bound on 
$m_s$ to be taken seriously, the corresponding decay signature should
be large compared to any such features.  
In practice, the limits we determine with this method roughly correspond to the lowest mass for which the 
decay signal more than doubles the astrophysical background in a particular energy bin.
In other words, to invalidate our upper bound the entire astrophysical background flux would
have to vanish \it{only}\rm~in the bin in which we set our limit. 

In the case of Andromeda, the subtraction of comparable total and discrete point source emission within the 5$'$ extraction region, $L_{\rm x,TOT} = (2.2\pm 0.2) \times 10^{39}~\rm{erg}~\rm{s}^{-1}$ vs. $L_{\rm x,PT} = (2.0\pm 0.1) \times 10^{39}~\rm{erg}~\rm{s}^{-1}$, leads to a factor of two uncertainty in the normalization of the diffuse spectrum:
$L_{\rm x,DIFFUSE} \simeq (0.2\pm 0.2)\times 10^{39}~\rm{erg}~\rm{s}^{-1}$ \cite{M31}.  To account for this,
we require the decay peak associated with the Andromeda mass limit to be at least four times the astrophysical background in the relevant energy bin. Based on this even more stringent criterion,
i.e., 95\% C.L.$\equiv 4\sigma_{\rm f}$,
the diffuse spectrum of Andromeda enables us to set 
an upper bound of $m_s < 3.5$ keV (95\% C.L.).
Because of the rapidly increasing decay signal ($\propto m_s^{3.374}$)
and rapidly falling background, marginally larger
values of $m_s$ are excluded at even higher confidence levels.  We note that our limit
is significantly more restrictive than the upper bound determined, through a similarly conservative
process, from the spectrum of Virgo A ($m_s < 8.2$ keV, 95\% C.L.) \cite{Aba05b}.
The decay signatures corresponding to
both of these limits are shown in Fig.~\ref{Galspec}. For comparison, we also show 
the decay signatures that would be present if Andromeda's dark matter halo
were composed of sterile neutrinos with a mass
of $m_s = 6.3$ keV (the estimated Virgo A + Coma limit \cite{Boyarsky2006,AbaSav}, see Sec. I) or $m_s = 8.2$ keV (the Virgo A limit \cite{Aba05b}).
The decay peak corresponding to 8.2 keV sterile neutrinos
has a much more pronounced
appearance in Andromeda than in Virgo A for two reasons:
(1) the measured (background) spectrum of Andromeda is almost two orders of
magnitude lower than that of Virgo A, yet,
(2) regardless of the value of $m_s$,
$\Phi^{\rm M31}_{\rm x,s}$ is comparable to $\Phi^{\rm M87}_{\rm x,s}$
(Eqn.~\ref{Phixs_nomix}): 
\beq
\frac{\Phi^{\rm M31}_{\rm x,s}}{\Phi^{\rm M87}_{\rm x,s}} = 
\frac{D^2_{\rm M87}}{D^2_{\rm M31}}
\frac{M^{\rm fov}_{\rm DM, M31}} {M^{\rm fov}_{\rm DM,M87}} 
\simeq  0.71.
\label{DMflux}
\eeq
If the backgrounds of Andromeda and Virgo A were energy-independent, we would expect to be able to
improve the Virgo A limit by a factor of $\simeq 100^{0.3} \simeq 4$ (Eqn.~\ref{Phixs_nomix}),
to $m_s ~\lsim 2$ keV.  However, because the X-ray spectrum of Andromeda rapidly increases at low energies, as galactic and cluster X-ray spectra typically do, we do not quite reach this rough expectation.

\begin{figure}
\includegraphics[scale=0.425]{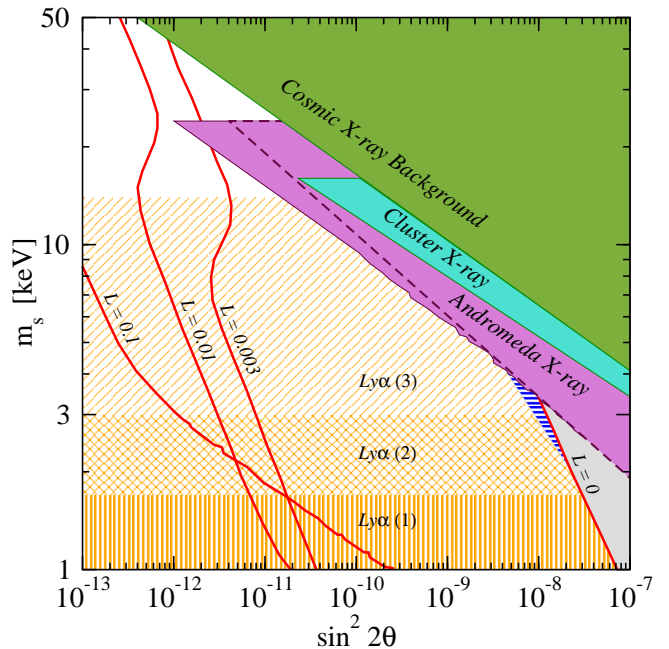}
\caption {Here we present constraints on $m_s$ as a function of 
mixing angle, sin$^2 2\theta$, assuming that all dark matter is comprised of sterile
neutrinos. 
To facilitate comparisons, we adopt many of the conventions used by Abazajian
and Koushiappas \cite{AbaSav}. 
For $L = 0$, the thick, solid line corresponds to $\Omega_s= 0.24$ (Eqn.~\ref{Omega_sin2th}),
while the shaded region to the right corresponds to $\Omega_s > 0.24$.
Three density-production relationships associated
with $\Omega_s= 0.3$ and $L \gg 10^{-10}$ are also shown \cite{AbaSav}. 
The two previous \it{direct}\rm~radiative decay ($\nu_s \rightarrow \nu_{e,\mu,\tau} + \gamma$)
upper limits (both 95\% C.L.) are based on measurements \cite{Lumb,ReadPonX,Marshall,Gruber} of the 
Cosmic X-ray Background~\cite{Boyarsky:2005us} and 
\it{XMM}\rm~observations \cite{VirgoX,Briel2001,Neumann2003} of Virgo A (M87) and the Coma
cluster \cite{Aba05b,Boyarsky2006,AbaSav}.
The most stringent direct limits, from the present work (also 95\% C.L.),
are based on \it{XMM}\rm~observations of the
Andromeda galaxy \cite{M31}. The region bounded by the dashed line is excluded by the
``$\Gamma_{\rm s,tot}-$scaling method'', while
the region above the solid, slightly jagged line is excluded by the more accurate 
``direct data method'' (see Sec.~\ref{mslimit}). 
The \it{indirect}\rm~lower limits (all 95\% C.L.) labeled Ly$\alpha$(1) and Ly$\alpha$(2)
were derived in Ref.~\cite{Aba05b}, while Ly$\alpha$(3) was derived
in Ref.~\cite{Seljak}.  Sterile neutrinos that occupy the horizontally
hatched region could explain pulsar kicks \cite{Kusenko1998,Fuller2003,Kusenko2004,Barkovich2004,Fryer2005}.
\label{XMM_ms}}
\end{figure}

\section{Exclusion Limits in the $m_{s~} -$~$\rm{sin}^{2}2\theta$ plane}
\label{mslimit}

To determine the region of the $m_{s~} -$~sin$^{2}2\theta$ plane (Fig.~\ref{XMM_ms}) 
that is excluded by the X-ray spectra of Andromeda and Virgo A, we must generalize the mass limits
shown in Fig.~\ref{Galspec} (which were derived under the assumption that $\Omega_s = \Omega_{\rm DM} = 0.24$)
to arbitrary values of $\Omega_s$.  
One approach is to derive a smooth exclusion
region by scaling the total rate of $\nu_s$ radiative decays from a halo of mass
$M_{\rm DM}$,
\beq
\Gamma_{\rm s,tot} = \left(\frac{M_{\rm DM}}{m_s}\right)\Gamma_s 
\propto m_s^4 \rm{sin}^{2}2\theta.
\label{Fdiff}
\eeq
The key to this ``$\Gamma_{\rm s,tot}-$scaling method'', which is used in \cite{AFT,Aba05a,Aba05b},
is to base the scaling on a $\nu_s$ decay peak (evaluated at $\Omega_s$ = 0.24)
that is large compared to both the formal statistical errors on the flux as well as 
any fluctuations in the spectral data, as we have done.

For each mass limit associated with such a peak, $m_{s,\rm lim}$,
we determine the mixing angle, sin$^{2}2\theta_{\Omega_s}$ (via Eqn.~\ref{Omega_sin2th}),
that corresponds to $\Omega_s$ = $\Omega_{\rm DM}$ = 0.24.  
By restricting alternative values of $m_s$ and sin$^{2}2\theta$ so that
$\Gamma_{\rm s,tot}$ will remain fixed at 
$\Gamma_{\rm s,tot}(m_{s,\rm lim},\rm{sin}^{2}2\theta_{\Omega_s})$, we arrive at the following scaling
of each mass limit through the $m_{s~} -$~sin$^{2}2\theta$ plane: 
\beq
m_s ~\lsim  m_{s,\rm lim}
\left(\frac{\rm{sin}^{2}2\theta}{\rm{sin}^{2}2\theta_{\Omega_s}}\right)^{-1/4}.
\label{mslimtheta}
\eeq
In the case of Andromeda, we find
\beqa
m_s ~\lsim  3.5 \rm~keV \left(\frac{\rm{sin}^{2}2\theta}{8.94\times 10^{-9}}\right)^{-1/4} \nonumber \\
= 1.91 \rm~keV \left(\frac{\rm{sin}^{2}2\theta}{10^{-7}}\right)^{-1/4},
\label{mslimtheta_M31}
\eeqa
which excludes the region bounded by the dashed line (and above) in Fig.~\ref{XMM_ms}.

There is an alternative and more accurate approach \cite{Boyarsky2006} through which a limit on $m_s$ can be established
more directly from the spectral data.
For each value of $m_s$, one
determines the value of sin$^{2}2\theta$ for which
$\Phi_{\rm x,s} (m_s , \rm{sin}^2 2\theta)$ (Eqn.~\ref{Phixs}) exceeds the 
measured spectrum at $ E_{\gamma ,\rm s} = m_s/2$ by some threshold ($4 \sigma_{\rm f}$ in our case).
We apply this ``direct data method'' to the (0.5$-$12 keV) diffuse spectrum of Andromeda.
Above 5 keV, where the spectrum is dominated by instrumental background, we fix the amplitude
at 0.003 Counts/sec/keV, which is well above the steeply falling data \cite{M31}.
The conservative (95\% C.L.) exclusion region that results is bounded by the slightly jagged line in Fig.~\ref{XMM_ms}.
An approximate fit to the boundary of this region is
\beq
m_s ~\lsim  2.1 \rm~keV \left(\frac{\rm{sin}^{2}2\theta}{10^{-7}}\right)^{-0.213},
\label{mslimtheta_M31B}
\eeq
which is roughly parallel to the boundaries of the Cluster \cite{Boyarsky2006,AbaSav} and CXB
\cite{Boyarsky:2005us} regions that were derived using essentially the same method
(see Fig.~\ref{XMM_ms} and Eqns.~\ref{msCXB} and \ref{msComa} below). 

In addition to 
our Andromeda bounds, 
two previous radiative decay limits are also shown in 
Fig.~\ref{XMM_ms}.  
Boyarsky et al.~\cite{Boyarsky:2005us} constrained $m_s$ using
\it{XMM-Newton}\rm~(0.5$-$12 keV) \cite{Lumb,ReadPonX}
and HEAO-I (3$-$60 keV) \cite{Marshall,Gruber}
measurements of the CXB.
The resulting limit ($m_s <$ 9.3 keV at $\Omega_s = 0.24$) follows the trend~\cite{Boyarsky:2005us}:
\beq
m_s ~ \lsim 4.1 \rm~keV \left(\frac{\rm{sin}^{2}2\theta}{10^{-7}}\right)^{-1/5}
\left(\frac{\Omega_{\rm DM}}{0.24}\right)^{-1/5}.
\label{msCXB}
\eeq 
According to the analyses in Refs. \cite{Boyarsky2006, AbaSav} the
\it{XMM}\rm~observations
of Virgo A (M87) \cite{VirgoX} and the Coma cluster \cite{Briel2001,Neumann2003} demand:
\beq
m_s ~ \lsim 3.4 \rm~keV \left(\frac{\rm{sin}^{2}2\theta}{10^{-7}}\right)^{-0.184}.
\label{msComa}
\eeq

Boyarsky et al. \cite{Boyarsky2006b} recently determined a limit on sterile neutrino
decays based on \it{XMM}\rm~observations of the Large Magellanic Cloud (LMC).  
We note that the signals shown in their spectra are very weak compared to backgrounds.
Nevertheless, their claimed exclusion region is less restrictive than ours.
Abazajian and Koushiappas \cite{AbaSav} have also questioned the robustness of the LMC limit and all
constraints based on dwarf/satellite galaxies due to the large uncertainties in their dark matter distributions. 

The indirect constraints found in Ref.~\cite{Aba05b}, which we have also reproduced
in Fig.~\ref{XMM_ms}, were derived using  
small scale clustering data from a variety of CMB 
measurements (WMAP~\cite{hinshaw03,kogut03}, CBI~\cite{readhead04}, 
Boomerang~\cite{Jones:2005yb}, ACBAR~\cite{Kuo:2002ua}, VSA~\cite{dickinson04}), 
the SDSS 3-D galaxy power
spectrum, $P_g(k)$ \cite{Tegmark:2003uf}, the linear matter power
spectrum inferred from Ly-$\alpha$
absorption in the SDSS quasar
catalog~\cite{McDonald:2004eu,McDonald:2004xn} and from high-resolution observations 
of the Ly-$\alpha$ forest \cite{Croft:2000hs,Viel:2004bf}.
When combined, the CMB measurements,
SDSS 3-D $P_g(k)$, and SDSS Ly-$\alpha$ forest observations set a lower
bound (Ly$\alpha$(1)) of \cite{Aba05b}:
\beq
m_s > 1.7\rm\ keV~~~(CMB+P_g(k)+Ly\rm{-}\alpha ~~ 95\% ~C.L.).
\label{msCMB_SDSS}
\eeq

The high-resolution Ly-$\alpha$ forest data could potentially provide
an even stronger lower limit on $m_s$, were it not for the
($15\% - 30\%$) systematic errors \cite{Viel:2004np,Viel:2005qj,Aba05b}.
In the limit of 15\% (Gaussian) systematic \cite{Croft:2000hs,Viel:2004bf} uncertainties, combining 
the high-resolution data of \cite{Viel:2004bf} with the three
data sets used to derive Eqn.~(\ref{msCMB_SDSS}) yields (Ly$\alpha$(2)) \cite{Aba05b}:
\beq
m_s > 3.0\rm\ keV~~~(\rm Previous + HR~ Ly\rm{-}\alpha ~~ 95\% ~C.L.).
\label{msCMB_SDSS_HRLya}
\eeq

\noindent Recently, Seljak et al. \cite{Seljak} recalculated this lower bound, finding
$m_s > 14$ keV (95\% C.L.), which we also show in Fig.~\ref{XMM_ms} (Ly$\alpha$(3)).
As discussed in the introduction, these results are still quite new and the subject
of some controversy \cite{Aba05b, AbaSav}, but
the basic agreement between these findings
and the independent analysis of Viel et al. \cite{Viel2006} ($m_s > 10 \rm~keV$ 95\% C.L.), 
strengthens the case for such a restrictive, indirect limit.
In any case, it is valuable to improve both the direct
and indirect constraints separately.

\section{Conclusions}
\label{conclusions}

Warm Dark Matter (WDM) models of structure formation 
may more easily explain the low level of observed small scale clustering
than standard $\Lambda$CDM galaxy formation scenarios.  In this paper,
we used the diffuse X-ray flux of the Andromeda galaxy (M31) \cite{M31}
to improve the radiative decay ($\nu_s \rightarrow \nu_{\rm e,\mu ,\tau} + \gamma$)
upper limits on the mass $m_s$ of sterile neutrino WDM.
In the context of the model described in \cite{AFP,AFT,Aba05a,Aba05b},
our analysis of the diffuse spectrum of Andromeda requires $m_s < 3.5 \rm\ keV$ (95$\%$ C.L.).  
Because of the rapidly increasing sterile neutrino decay signal ($\propto m_s^{3.374}$)
and rapidly falling background, larger
values of $m_s$ are excluded at much higher confidence levels.
As a case in point, we demonstrated that the decay signature associated
with the upper limit ($m_s < 8.2$ keV; 95$\%$ C.L. \cite{Aba05b}) 
inferred from the spectrum of Virgo A (M87)
would be enormous relative to the significantly reduced astrophysical background  
of Andromeda. 
When we combine our \it{direct}\rm~constraint with the most conservative,
\it{indirect}\rm~lower limit set by 
measurements of small scale clustering in the CMB,
the SDSS 3-D galaxy power spectrum, and the Ly-$\alpha$ forest \cite{Aba05b},
we find that $m_s$ is restricted to the narrow range
\begin{equation}
1.7{\rm\ keV} < m_s < 3.5\rm\ keV ~~~(95\% ~\rm C.L.),
\end{equation}
in the $L\simeq 10^{-10} \simeq 0$ case.  If the sterile neutrinos occupy this tiny window,
they could still be the dark matter and generate pulsar kicks \cite{Kusenko1998,Fuller2003,Kusenko2004,Barkovich2004,Fryer2005},
but most of the parameter space remains viable
only if the lepton asymmetry is very large: $L \gg 10^{-10}$. 
Indeed the corroboration of the Seljak et al. ($m_s > 14$ keV; 95$\%$ C.L.) and Viel et al. ($m_s > 10$ keV; 95$\%$ C.L.)
Ly$\alpha$ constraints \cite{Seljak, Viel2006} strongly suggest that the standard $L=0$ production scenario of Abazajian et al.
\cite{AFT,AFP,Aba05a} is no longer viable.

A point source-subtracted \it{XMM}\rm~and/or \it{Chandra}\rm~spectrum
of Andromeda (and/or other nearby, massive yet quiescent spiral galaxies)
extracted from within a radius of
$\geq 50' - 500'$ would probe $\gsim$ 10$-$100 times more dark matter than the
observations we have considered here \cite{M31}.  For a sufficiently large extraction
region, the enhancement in the predicted $\nu_s$ decay signal should be large enough
compared to the additional diffuse, astrophysical background enclosed
to provide sensitivity to sterile neutrino masses as low as $m_s$ = 1.7 keV.
Detailed modeling of
astrophysical emission and instrumental response, which we have not attempted in the present paper, might
enable us to set even more stringent constraints by allowing us
to relax our limiting criterion without sacrificing robustness.
In addition to providing an even more restrictive bound on $m_s$, this initiative would
establish a much more comprehensive picture of the X-ray point source population and
diffuse emission of Andromeda (and/or other nearby spirals). The proposed observation(s)
would therefore greatly benefit both the X-ray astronomy and astroparticle physics and
cosmology communities.

At the same time, analyses of new and existing small scale structure data
and the cutoff scale in the linear matter power spectrum that they do (or do not) reveal
will provide steadily improving lower bounds on $m_s$.
The convergence of these complementary constraints should lead to 
the detection or exclusion of sterile neutrino WDM in the near term.

\acknowledgments
We thank Kev Abazajian, Alexei Boyarsky, Steen Hansen, Matthew Kistler,
Christopher Kochanek, Savvas Koushiappas, Alexander Kusenko, Roberto Soria, Louie Strigari, and Matteo Viel
for helpful discussions.  CRW acknowledges support from The Ohio State University 
Department of Physics and Department of Astronomy, Org. 06231.
JFB and HY acknowledge support from The Ohio State University and the
NSF CAREER grant No. PHY-0547102 to JFB. TPW
acknowledges support from The Ohio State University and Department of Energy grant No. DE-FG02-91ER40690.

\end{document}